\documentclass[12pt]{article}\usepackage{graphics}
\begin{document}
\title{Mass and flavor from strong interactions}
\author{\normalsize B. Holdom
\thanks{Talk given at the workshop on Fermion Mass and CP violation,
Hiroshima, Japan, March 1998.} \\\small {\em Department of Physics,
University of Toronto}\\\small {\em Toronto, Ontario,} M5S1A7,
CANADA}\date{}\maketitle
\begin{picture}(0,0)(0,0)
\put(310,205){UTPT-98-07}
\put(310,190){hep-ph/9804312}
\end{picture}

\begin{abstract} We provide an economical description of mass and flavor
based on strong interactions and some dynamical assumptions. We include a
discussion of CP violation in the quark sector and its relation to neutrino
masses.
\end{abstract}
\baselineskip 19pt

\section{Introduction} Over the years there have two basic approaches to
finding the theory of flavor and mass.\footnote{Theories which simply
parametrize fermion masses, such as the standard model, do not qualify as
theories of flavor and mass.} One approach is to consider only those theories
we fully understand. That is, one constructs theories which either are
perturbative, or are based on some rigorous results of strong interactions, such
as those emerging in the study of strongly interacting supersymmetric
theories. But so far at least, this approach has led to models of mass and flavor
which appear overly complicated, with numerous new interactions and/or
matter multiplets. The other approach is to consider only those theories with
an economical structure. But so far at least, this approach has forced the
model builder to make dynamical assumptions about the behavior of strong
interactions. In other words, it is not known that the models being proposed
actually work as advertised.

It is not surprising that the first approach has proven much more popular over
the years; it is preferable to know what one is talking about! On the other
hand, the theory of hadronic interactions, QCD, has more in common with the
second approach. QCD has a simple and economical structure, and yet it is
often difficult to extract physical results. QCD is the theory of the hadronic
mass spectrum, but we have yet to see this spectrum fully emerge from a
theoretical calculation. Even the concept of confinement is still closer to a
dynamical assumption than a rigorously derived result of the theory. But none
of this shakes our acceptance of QCD, since there have been other ways to get
a handle on QCD which have allowed for experimental checks and
confirmation of the theory.

Our experience with QCD thus suggests that it is not necessarily fatal for a
theory to rely on plausible dynamical assumptions, as long as the structure of
the theory is rigid enough to lead to testable consequences. The correct theory
of flavor and mass may be such as to
\textit{not} allow for a calculation of the fermion mass spectrum with current
tools. But even without being able to provide precise numbers, the following
are examples of what we may hope to glean from the correct theory.

\begin{itemize}
\item  patterns in new flavor dependent effects
\item  patterns in CP violation
\item  patterns in neutrino masses
\item  predictions for the lightest of the new particles
\end{itemize}

Another common theme in the search for the theory of mass and flavor is to
first deal with the question of electroweak symmetry breaking. There are two
widely reported approaches to that question, supersymmetry and technicolor,
which both provide attractive answers to that single question. These are then
taken as the two possible starting points in the search for the theory of mass
and flavor. But as indicated in Fig.\ 1, many obstacles must be overcome in
each case before one can approach a comprehensive theory of mass and
flavor. In both cases, after the various hurdles are passed, the resulting
proposed theories are looking quite complicated and convoluted.

Here we shall consider the possibility that the key to electroweak symmetry
breaking is neither supersymmetry or conventional technicolor. We will hope
to identify an alternative which leads more simply and naturally to a theory of
flavor. The price we will pay is to have electroweak symmetry breaking
associated with some aspect of strong interactions which is less familiar to us,
i.e.\ associated with a dynamical assumption.

Our basic picture is as follows.

\begin{itemize}
\item  A new strong interaction breaks close to a TeV, unlike technicolor
which remains unbroken.
\end{itemize}
\begin{itemize}
\item  Associated with this symmetry breaking are the dynamically generated
masses for a fourth family of quarks and leptons, which in turn is responsible
for electroweak symmetry breaking.
\end{itemize}
\begin{itemize}
\item  The new strong interaction is a remnant flavor interaction, and it only
acts on the third and fourth families.
\item  At a higher ``flavor scale'', say 100--1000 TeV, the remnant flavor
interaction merges with the full flavor interaction, which involves all quarks
and leptons.
\end{itemize} The full flavor interaction is some strong, chiral gauge
interactions which partially breaks itself. The important point is that this
symmetry breaking does not include the breakdown of
\({\mathit{SU}(2)_{L}}\times {\mathit{U}(1)_{Y}}\), and thus the known
quarks and leptons receive no mass at the flavor scale. The exception are the
right-handed neutrinos, which can serve as bilinear order parameters for the
flavor breaking. The theory above the flavor scale may also be left-right
symmetric, in which case the right-handed neutrino condensates also serve to
break \({\mathit{SU}(2)_{L}}\times {\mathit{SU}(2)_{R}}\times
{\mathit{U}(1) _{B - L}}\) down to \({\mathit{SU}(2)_{L}}\times
{\mathit{U}(1)_{Y}}\).

The basic two-scale structure of the model is shown in Fig.\ 2. The physics at
the flavor scale shows up on lower scales through 4-fermion operators and
other nonrenormalizable operators. These effects, combined with the mass
generation at a TeV, feed down masses from the fourth family to the ligher
families. The following are some key ingredients for understanding the origin
of a complicated fermion mass spectrum.

\begin{itemize}
\item  There are a wide variety of possible 4-fermion operators, due to
strong-coupled flavor physics, and different operators can contribute to
different elements of the mass matrices.
\item  Operators have various transformation properties under the remnant
flavor interaction, and in particular, operators have various numbers of
fermions coupling to this interaction.
\end{itemize}
\begin{itemize}
\item  Since the remnant flavor interaction is strong, we can expect that
anomalous scaling gives large relative enhancement of operators. This will be
one of the sources of quark and lepton mass hierarchies.
\end{itemize}

One problem which has plagued the technicolor approach has been the
difficulty in understanding the origin of the large isospin breaking inherent in
the top mass, in a way compatible with the electroweak correction parameter
\(\delta \rho \). In our approach, isospin breaking originates at the flavor scale,
for example through a dynamical breakdown of \({\mathit{SU}(2)_{R}}\).
The remnant flavor interaction remaining down to a TeV is isospin preserving
and it is this interaction which is responsible for electroweak symmetry
breaking. This in itself produces no contribution to \(\delta \rho \). Isospin
breaking is communicated to the TeV scale via 4-fermion operators, and an
operator in particular which must be present is the \(t\)-mass operator
\(\overline{\mathit{t'}}\mathit{t'}
\overline{t}t\), where primes denote fourth family members. (It may be that
the corresponding \(b\)-mass operator, \(\overline{\mathit{b'}}\mathit{b'}
\overline{b}b\), is generated as a weak radiative \({\mathit{SU}(2)_{R}}\)
correction to the \(t\)-mass operator.) It turns out that the contribution of the
\(\overline{\mathit{t'}}\mathit{t'}
\overline{t}t\) operator to \(\delta \rho \) is suppressed by
\(({m_{t}}/{m_{\mathit{t'}}})^{4}\) where \({m_{\mathit{t'}}}\approx
1\mathrm{\ TeV}\), and thus the \(t\)-mass does not directly imply a
significant problem for \(\delta \rho \). Indeed we are relying on how small the
\(t\) mass is small relative to the fundamental TeV scale, which differs from
the usual emphasis on how large the \(t\) mass is. We shall find a dynamical
reason as to why the \(t\)-mass operator is the largest isospin violating
operator, due to the anomalous scaling mentioned above.

\section{A minimal model} We will now specify the model in more detail
\cite{a}. The main object here is to show how a complicated fermion mass
spectrum can arise from a simple underlying structure. It is sufficient for us to
present the minimal model, since we do not have adequate understanding of
the strong dynamics to judge which variation of the model will produce the
assumed symmetry breaking pattern.\footnote{But see \cite{b}. Less minimal
versions will likely have a new sector of strongly interacting fermions which
play little role in quark and lepton mass generation. The most minimal model
would seem to be preferred by the constraints on \(S\) and \(T\), but see
\cite{c}.} We will consider a 4 family model where the flavor gauge
symmetry is \({\mathit{U}(2)_{V}}\).\footnote{The subscript reminds us
that this is a vectorial interaction in a certain basis. An additional axial
interaction at the flavor scale, which plays little role in our discussion, is
needed to make the strong interaction chiral.} Two pairs of families transform
as \((2, +)\) and \((\overline{2}, -)\) under
\({\mathit{SU}(2)_{V}}{\times}{\mathit{U}(1) _{V}}\); we label these two
pairs of families as \([Q, L]\) and \([\underline{Q}, \underline{L}]\). The
basic structure of the model, including the right-handed neutrino masses
which are assumed to occur at the flavor scale and the resulting breakdown of
\({\mathit{U}(2)_{V}}\) to \({\mathit{U}(1)_{X}}\), are depicted in Fig.\ 3.
Notice that \({\mathit{U}(1)_{X}}\) couples only to the two heavy families,
and that the fermion basis depicted in the figure is not the mass eigenstate
basis.

The main dynamical assumption we make is in the form of the fourth family
masses. These masses must be generated by the strong, and broken,
\({\mathit{U}(1)_{X}}\) interaction along with possible 4-fermion
interactions.\footnote{The fact that the 4th family and not the 3rd family
masses form must be due to a cross-channel coupling, which should be
familiar to builders of multi-Higgs potentials. Note that the \(\tau '\) mass
forms in the \({\mathit{U}(1)_{X}}\)-singlet channel, unlike the \(q'\)
masses, which may be due to flavor induced 4-fermion operators which
distinguish quarks and leptons and which are enhanced by
\({\mathit{U}(1)_{X}}\) anomalous scaling, e.g. \(\overline{{\tau
_{L}}'}{\tau _{R}}'
\overline{{\tau _{R}}'}{\tau _{L}}'\).}

\begin{itemize}
\item  \(\mathit{t'}\) and \(\mathit{b'}\) quark masses:
\({\overline{\underline{Q}}_{\mathit{L1}}}{Q_{\mathit{R1}}}\)
\item  \(\tau '\) mass:
\({\overline{\underline{E}}_{\mathit{L1}}}{\underline{E}_{\mathit{R1}}}\)
\item  \({\nu _{\tau '}}\) mass: \({\underline{N}_{\mathit{L1}}^2}\)
\end{itemize} Now consider the 4-fermion operators which feed these masses
to the lighter quark and leptons. We find that interesting results follow from
the following subset of operators. Other operators may also contribute, but we
assume that this subset dominates. The unique characteristics of these
operators can of course provide a dynamical reason for their dominance.

\begin{itemize}
\item  They have the chiral structure \({\overline{\psi }_{L}}{\psi
_{R}}{\overline{\psi }_{L}}{\psi _{R}}\), and hence must be generated
dynamically by the strong flavor interactions.
\end{itemize}
\begin{itemize}
\item  They preserve \({\mathit{SU}(2)_{L}}{\times}{\mathit{U
}(1)_{Y}}\) but display maximal \({\mathit{SU}(2)_{R}}\) breaking.
\item  They preserve the strong \({\mathit{SU}(2)_{V}}\). They may thus be
composed of the \({\mathit{SU}(2)_{V}}\) singlet bilinears: 
\({\overline{Q}_{\mathit{Li}}}{Q_{\mathit{Ri}}}\),
\({\overline{Q}_{\mathit{Li}}}{\underline{Q}_{\mathit{Rj}}}{\varepsilon
_{\mathit{ij}}}\), etc.
\end{itemize} Some of these operators break \({\mathit{U}(1)_{X}}\), and
we will assume that this generates an \(X\) mass of order a TeV, or somewhat
higher.

\section{Quark masses} We now briefly describe the various operators which
are responsible for quark and lepton masses. We first discuss the quark sector.
The following operators feed mass down from \(\mathit{t'}\) and
\(\mathit{b'}\) (and in the last case from \(t\)):
\begin{equation}
 \left[  {\begin{array}{cc}
{\overline{U}_{\mathit{L1}}}{\mathit{D}_{\mathit{R1}}}{\overline{\underline{D}}_{\mathit{L1}}}{\underline{U}_{\mathit{R1
}}} & {\cal{B}} \\
{\overline{\mathit{D}}_{\mathit{L1}}}{U_{\mathit{R1}}}{\overline{\underline{U}}_{\mathit{L1}}}{\underline{D}_{\mathit{R1
}}} & {\tilde{\cal{B}}} \\
{\overline{U}_{\mathit{L1}}}{\mathit{D}_{\mathit{R1}}}{\overline{\underline{D}}_{\mathit{L1}}}{U_{\mathit{R2}}}
&  {\cal{C}} \\
{\overline{\mathit{D}}_{\mathit{L1}}}{U_{\mathit{R1}}}{\overline{\underline{U}}_{\mathit{L1}}}{\mathit{D}_{\mathit{R2}}
} & {\tilde{\cal{C}}} \\
{\overline{\underline{U}}_{\mathit{L2}}}{\mathit{D}_{\mathit{R1}
}}{\overline{\underline{D}}_{\mathit{L1}}}{\underline{U}_{\mathit{R1}}}
& {\cal{D}} \\
{\overline{\underline{D}}_{\mathit{L2}}}{U_{\mathit{R1}}}{\overline{\underline{U}}_{\mathit{L1}}}{\underline{D}_{\mathit{R1
}}} & {\tilde{\cal{D}}} \\
{\overline{\underline{Q}}_{\mathit{Li}}}{U_{\mathit{Rj}}}{\varepsilon
_{\mathit{ij}}}{\overline{\underline{Q}}_{\mathit{Lk
}}}{\mathit{D}_{\mathit{Rl}}}{\varepsilon _{\mathit{kl}}} &  {\cal{E}}
\\ {\overline{Q}_{\mathit{Li}}}{\underline{U}_{\mathit{Rj}}}{\varepsilon
_{\mathit{ij}}}{\overline{Q}_{\mathit{Lk}}}{\underline{D}_{\mathit{Rl}}}{\varepsilon
_{\mathit{kl}}} &  {\cal{F}}
\end{array}}
 \right] \end{equation} while the following operators feed mass down from
\(\tau '\):
\begin{equation}
 \left[  {\begin{array}{cc}
{\overline{\underline{E}}_{\mathit{L1}}}{\underline{E}_{\mathit{R1}}}{\overline{U}_{\mathit{L1}}}{U_{\mathit{R1}}}
& {{\cal{G}}_{1}} \\
{\overline{\underline{E}}_{\mathit{L1}}}{\underline{E}_{\mathit{R1}}}{\overline{U}_{\mathit{L2}}}{U_{\mathit{R2}}}
& {{\cal{G}}_{2}} \\
{\overline{\underline{E}}_{\mathit{L1}}}{\underline{E}_{\mathit{R1}}}{\overline{\underline{U}}_{\mathit{L1}}}{\underline{U}_{\mathit{R1}}}
& {{\cal{H}}_{1}} \\
{\overline{\underline{E}}_{\mathit{L1}}}{\underline{E}_{\mathit{R1}}}{\overline{\underline{U}}_{\mathit{L2}}}{\underline{U}_{\mathit{R2}}}
& {{\cal{H}}_{2}} \\
{\overline{\underline{E}}_{\mathit{L1}}}{\underline{E}_{\mathit{R1}}}{\overline{U}_{\mathit{Li}}}{\underline{U}_{\mathit{Rj}}}
{\varepsilon _{\mathit{ij}}} & {\cal{I}} \\
{\overline{\underline{E}}_{\mathit{L1}}}{\underline{E}_{\mathit{R1}}}{\overline{\underline{U}}_{\mathit{Li}}}{U_{\mathit{Rj}}}
{\varepsilon _{\mathit{ij}}} & {\cal{J}}
\end{array}}
 \right] \end{equation} The main point is that each operator contributes to a
different mass element.
\begin{equation} {M_{u}}= \left[  {\begin{array}{cccc} 0 &
{{\cal{G}}_{2}} & {\cal{I}} & 0 \\ {{\cal{H}}_{2}} & {\cal{E}} &
{\cal{D}} & {\cal{J}}
 \\ {\cal{I}} & {\cal{C}} & {\cal{B}} & {{\cal{G}}_{1}}
 \\ 0 & {\cal{J}} & {{\cal{H}}_{1}} & {\cal{A}}
\end{array}}
 \right] \end{equation}
\begin{equation} {M_{d}}= \left[  {\begin{array}{cccc} {\cal{F}} & 0 & 0
& 0 \\ 0 & {\cal{E}} & {\tilde{\cal{D}}} & 0 \\ 0 & {\tilde{\cal{C}}} &
{\tilde{\cal{B}}} & 0 \\ 0 & 0 & 0 & {\cal{A}}
\end{array}}
 \right] \end{equation} Note that essentially all of the CKM mixing arises in
the up sector, and that the mass matrices are not symmetric. Various mass
hierarchies arise for the following reasons.

\begin{itemize}
\item  Various operators experience different power-law scaling
enhancements from the strong \({\mathit{U}(1)_{X}}\). Basically, operators
containing heavy fermions in both Lorentz and \({\mathit{U}(1)_{X}}\)
singlet combinations are expected to be enhanced the most. The \({\cal{B}}\)
operator, which is the \(t\)-mass operator, is expected to be the largest.
\begin{eqnarray}&&{\cal{B}}>{\cal{C}},
{\cal{D}}>{\cal{E}}\\&&{{\cal{G}}_{1}}, {{\cal{H}}_{1}}>{\cal{I}}, 
{\cal{J}}>{{\cal{G}}_{2}}, {{\cal{H}}_{2}}\end{eqnarray}
\item  There are different heavy masses, \({m_{\mathit{t',b'}}}>{m_{\tau ^{'
}}}>{m_{t}}\), being fed down.
\begin{eqnarray}&&{\cal{E}}>{\cal{F}}\\&&{\cal{B}}>{{\cal{G}}_{1}},
{{\cal{H}}_{1}}\\&&{\cal{C}}, {\cal{D}}>{\cal{I}},
{\cal{J}}\end{eqnarray}
\item  \({\tilde{\cal{B}}}\), \({\tilde{\cal{C}}}\) and \({\tilde{\cal{D}}}\)
can arise from weak radiative corrections (from \({\mathit{SU}(2)_{R}}\)).
\begin{equation}{\cal{B}}, {\cal{C}}, {\cal{D}}>{\tilde{\cal{B}}},
{\tilde{\cal{C}}},  {\tilde{\cal{D}}}\end{equation}
\item  Operators are affected differently by the axial interaction mentioned in
footnote 3.
\begin{equation}{\cal{G}}>{\cal{H}}\end{equation}
\end{itemize} We get one approximate relation due to the similarity of the
\({\cal{E}}\) and \({\cal{F}}\) operators.
\begin{equation}\frac {{m_{d}}}{{m_{s}}}\approx \frac
{{m_{t}}}{{m_{\mathit{t'}}}}\end{equation}
\section{Lepton masses} We now turn to the charged lepton mass matrices
where we find that the mixed quark-lepton operators play a crucial role. The
following operators feed mass down from \(\mathit{t'}\),
\begin{equation}
 \left[  {\begin{array}{cc}
{\overline{E}_{\mathit{L1}}}{U_{\mathit{R1}}}{\overline{\underline{U}}_{\mathit{L1}}}{\underline{E}_{\mathit{R1}}}
& {{\cal{B}}_{{\ell}}} \\
{\overline{E}_{\mathit{L1}}}{U_{\mathit{R1}}}{\overline{\underline{U}}_{\mathit{L1}}}{E_{\mathit{R2}}}
& {{\cal{C}}_{{\ell}}} \\
{\overline{\underline{E}}_{\mathit{L2}}}{U_{\mathit{R1}}}{\overline{\underline{U}}_{\mathit{L1}}}{\underline{E}_{\mathit{R1
}}} & {{\cal{D}}_{{\ell}}} \\
{\overline{\underline{E}}_{\mathit{L2}}}{U_{\mathit{R1}}}{\overline{\underline{U}}_{\mathit{L1}}}{E_{\mathit{R2}}}
& {{\cal{E}}_{{\ell}}}
\end{array}}
 \right] \end{equation} while the following operators feed mass down from
\(t\).
\begin{equation}
 \left[  {\begin{array}{cc}
{\overline{\underline{E}}_{\mathit{L1}}}{\underline{U}_{\mathit{R1}}}{\overline{U}_{\mathit{L1}}}{E_{\mathit{R1}}}
& {{\cal{F}}_{{\ell}}} \\
{\overline{E}_{\mathit{L2}}}{\underline{U}_{\mathit{R1}}}{\overline{U}_{\mathit{L1}}}{E_{\mathit{R1}}}
& {{\cal{G}}_{{\ell}}} \\
{\overline{\underline{E}}_{\mathit{L1}}}{\underline{U}_{\mathit{R1}}}{\overline{U}_{\mathit{L1}}}{\underline{E}_{\mathit{R2}}}
 & {{\cal{H}}_{{\ell}}} \\
{\overline{E}_{\mathit{L2}}}{\underline{U}_{\mathit{R1}}}{\overline{U}_{\mathit{L1}}}{\underline{E}_{\mathit{R2}}}
& {{\cal{I}}_{{\ell}}}
\end{array}}
 \right] \end{equation} The following operators are the only ones we mention
which are generated by \({\mathit{SU}(2)_{V}}\) exchange, and they feed
mass down from the \(\tau '\) and \(\tau \).
\begin{equation}
 \left[  {\begin{array}{cc}
{\overline{\underline{E}}_{\mathit{L1}}}{\underline{E}_{\mathit{R1}}}{\overline{\underline{E}}_{\mathit{R2}}}{\underline{E}_{\mathit{L2}}}
& {{\cal{J}}_{{\ell}}} \\
{\overline{E}_{\mathit{L1}}}{E_{\mathit{R1}}}{\overline{E}_{\mathit{R2}}}{E_{\mathit{L2}}}
& {{\cal{K}}_{{\ell}}}
\end{array}}
 \right] \end{equation} Here is the resulting matrix.
\begin{equation} {M_{{\ell}}}= \left[  {\begin{array}{cccc}
{{\cal{K}}_{{\ell}}} & {{\cal{I}}_{{\ell}}} & {{\cal{G}}_{{\ell}}} & 0 \\
{{\cal{E}}_{{\ell}}} & {{\cal{J}}_{{\ell}}} & 0 & {{\cal{D}}_{{\ell}}} \\
{{\cal{C}}_{{\ell}}} & 0 & 0 & {{\cal{B}} _{{\ell}}} \\ 0 &
{{\cal{H}}_{{\ell}}} & {{\cal{F}}_{{\ell}}} & {{\cal{A}}_{{\ell}}}
\end{array}}
 \right] \end{equation} The \(\mu \) mass is reasonable,
\begin{equation}{m_{\mu }}\approx \frac
{(1\mathrm{TeV})^{3}}{(100\mathrm{TeV}
)^{2}}{\label{mu}}\end{equation} and there is a relation due to the
similarity of the \({{\cal{J}}_{{\ell}}}\) and \({{\cal{K}}_{{\ell}}}\)
operators:
\begin{equation}\frac {{m_{e}}}{{m_{\mu }}}\approx \frac {{m_{\tau
}}}{{m_{\tau '}}}\end{equation}

We now turn to neutrinos. We have already mentioned that the RH neutrinos
have mass at the flavor scale, and that the 4th LH neutrino has a dynamical
mass in the 100 GeV to 1 TeV range. The remaining 3 LH neutrino masses
can only come from 6-fermion operators. For example, the operator
\begin{equation}{\overline{\underline{E}}_{\mathit{L1}}}{\underline{E}_{\mathit{R1}}}{\overline{\underline{E}}_{\mathit{L1}}}{\underline{E}_{\mathit{R1}}}{\overline{N}_{\mathit{L2}}}{\overline{N}_{\mathit{L2}}}\end{equation}
is generated by two
\({\overline{\underline{E}}_{\mathit{L1}}}{\underline{E}_{\mathit{R1}}}{\overline{N}_{\mathit{L2}}}{N_{\mathit{R2}}}\)
operators after integrating out the large \({N_{\mathit{R2}}}\) mass. The
result is that the \(\tau '\) mass feeds down to produce a small
\({N_{\mathit{L2}}}\) (i.e. \({\nu _{e}}\)) mass.\footnote{Using numbers
similar to those in (\ref{mu}), and accounting for the anomalous scaling
enhancement built into those numbers, can yield neutrino masses in the eV
range. We also note that in comparing these operators to those containing
quark fields, the latter operators may be dynamically favored due to QCD
effects.} This is very similar to the standard see-saw mechanism involving
scalar fields, except that the dimensions of the operators involved here are
much larger. This allows the right-handed neutrino mass scale to be at the
relatively low flavor scale we are discussing.

The whole set of 4-fermion operators which can contribute in this way to
neutrino masses are:
\begin{equation}
 \left[  {\begin{array}{cc}
{\overline{\underline{E}}_{\mathit{L1}}}{\underline{E}_{\mathit{R1}}}{\overline{N}_{\mathit{L2}}}{N_{\mathit{R2}}}
& {{\cal{B}}_{\nu }} \\
{\overline{\underline{E}}_{\mathit{L1}}}{\underline{E}_{\mathit{R1}}}{\overline{\underline{N}}_{\mathit{L2}}}{\underline{N}_{\mathit{R2}}}
& {{\cal{C}}_{\nu }} \\
{\overline{\underline{E}}_{\mathit{L1}}}{\underline{E}_{\mathit{R1}}}{\overline{N}_{\mathit{L1}}}{\underline{N}_{\mathit{R2}}}
 & {{\cal{D}}_{\nu }} \\
{\overline{\underline{E}}_{\mathit{L1}}}{\underline{E}_{\mathit{R1}}}{\overline{N}_{\mathit{L1}}}{N_{\mathit{R1}}}
& {{\cal{E}}_{\nu }} \\
{\overline{\underline{E}}_{\mathit{L1}}}{\underline{E}_{\mathit{R1}}}{\overline{\underline{N}}_{\mathit{L2}}}{N_{\mathit{R1}}}
 & {{\cal{F}}_{\nu }} \\
{\overline{\underline{E}}_{\mathit{L1}}}{\underline{E}_{\mathit{R1}}}{\overline{N}_{\mathit{L2}}}{\underline{N}_{\mathit{R1}}}
 & {{\cal{G}}_{\nu }}
\end{array}}
 \right] \end{equation} After labelling the heavy right-handed neutrino
masses as
\begin{equation}
 \left[  {\begin{array}{cc} {N_{\mathit{R2}}^2} & {m_{1}} \\
{\underline{N}_{\mathit{R2}}^2} & {m_{2}} \\
{N_{\mathit{R2}}}{\underline{N}_{\mathit{R2}}} & {m_{3}} \\
{N_{\mathit{R1}}}{\underline{N}_{\mathit{R1}}} & {m_{4}}
\end{array}}
 \right] \end{equation} we find for the light neutrino mass matrix the
following.
\begin{equation}
 \left[  {\begin{array}{ccc} {\displaystyle \frac {{{\cal{B}}_{\nu
}^2}}{{m_{1}}}}  &  {\displaystyle \frac {{{\cal{B}}_{\nu
}}{{\cal{C}}_{\nu } }}{{m_{3}}}}  + {\displaystyle \frac {{{\cal{F}}_{\nu
}}{{\cal{G}}_{\nu }}}{{m_{4}}}}  & {\displaystyle \frac
{{{\cal{B}}_{\nu }}{{\cal{D}}_{\nu }}}{{m_{3}}}}  +  {\displaystyle
\frac {{{\cal{E}}_{\nu }}{{\cal{G}}_{\nu } }}{{m_{4}}}}  \\[2ex]
{\displaystyle \frac {{{\cal{B}}_{\nu }}{{\cal{C}}_{\nu } }}{{m_{3}}}} 
+ {\displaystyle \frac {{{\cal{F}}_{\nu }}{{\cal{G}}_{\nu }}}{{m_{4}}}} 
& {\displaystyle \frac {{{\cal{C}}_{\nu }^2}}{{m_{2}}}}  & {\displaystyle
\frac {{{\cal{C}}_{\nu }}{{\cal{D}}_{\nu }}}{{m_{2}}}}  \\[2ex]
{\displaystyle \frac {{{\cal{B}}_{\nu }}{{\cal{D}}_{\nu } }}{{m_{3}}}} 
+ {\displaystyle \frac {{{\cal{E}}_{\nu }}{{\cal{G}}_{\nu }}}{{m_{4}}}} 
& {\displaystyle \frac {{{\cal{C}}_{\nu }}{{\cal{D}}_{\nu
}}}{{m_{2}}}}  &  {\displaystyle \frac {{{\cal{D}}_{\nu
}^2}}{{m_{2}}}} 
\end{array}}
 \right] \end{equation} We see that this matrix bears no resemblence to quark
or charged lepton mass matrices. Large mixings are expected, with masses
unrelated to the family hierarchy. CP violation is also expected, and it is to
that topic which we now turn.

\section{CP violation} We first note that the quantity most sensitive to
possible CP violation in new 4-fermion effects is the \(\varepsilon \)
parameter in the \(K\)-\(\overline{K}\) system. In other words we have a
natural setting for a superweak model of CP violation. Superweak models are
especially attractive in the context of the strong CP problem, since they allow
for the quark mass matrix to be real, or very close to it, which would account
for why the strong CP violating parameter \(\overline{\theta }\) is close to
vanishing. When we consider how this can arise in the present model, we find
that CP violation in the quark sector may arise in a way similar to neutrino
masses; that is, via 6-fermion operators only. This provides a natural
suppression mechanism which can go a long way towards suppressing strong
CP violation to acceptable levels.

In our picture we assume that above the flavor scale we have a CP invariant
gauge theory of massless fermions. Our dynamical assumption is that
lepton-number violation, \({\mathit{SU}(2)_{V}}\) breaking and CP
violation all originate in the right-handed neutrino condensates (both bilinear
and multilinear). We may then consider the operators which feed CP violation
feed into the quark sector. It can be shown that they must violate
lepton-number or \({\mathit{SU}(2)_{V}}\) or both. This in turn requires
6-fermion operators, of which the following are two examples.
\begin{eqnarray}&&{\overline{\mathit{D}}_{\mathit{L2}}}{\mathit{D}_{\mathit{R2}}
}{\overline{\mathit{D}}_{\mathit{L2}}}{\mathit{D}_{\mathit{R2
}}}{\overline{\underline{N}}_{\mathit{L1}}}{\overline{\underline{N}}_{\mathit{L1}}}\\&&{\overline{\underline{D}}_{\mathit{L2}}}{\underline{D}_{\mathit{R2}}}{\overline{\underline{D}}_{\mathit{L2}}}{\underline{D}_{\mathit{R2}}}{\overline{\underline{N}}_{\mathit{L1}}}{\overline{\underline{N}}_{\mathit{L1}}}\end{eqnarray}
From the mass matrices we have given it can be seen that in the presence of
the heavy \({\nu _{\tau '}}\) mass, these generate the \(\Delta S=2\) operators
\(({\overline{d}_{L}}{s_{R}})^{2}\) and
\(({\overline{s}_{L}}{d_{R}})^{2}\). If the 6-fermion operators have
coefficients of order \(1/(100\mathrm{\ TeV})^{5}\) and \(\langle
{\underline{N}_{\mathit{L1}}^2}\rangle \approx ( 1\mathrm{\
TeV})^{3}\), then the coefficients of the \(\Delta S=2\) operators are of the
right size to give \(\varepsilon \) in \(K\)--\(\overline{K}\) mixing.
\(\varepsilon '\) on the other hand requires \(d\)--\(s\) mass mixing, and thus is
negligible.

The following operators
\begin{eqnarray}&&{\overline{\mathit{D}}_{\mathit{L2}}}{\underline{D}_{\mathit{R1
}}}{\overline{\mathit{D}}_{\mathit{L2}}}{\underline{D}_{\mathit{R1}}}{\overline{\underline{N}}_{\mathit{L1}}}{\overline{\underline{N}}_{\mathit{L1}}}\\&&{\overline{\mathit{D}}_{\mathit{L1}}}{\underline{D}_{\mathit{R2
}}}{\overline{\mathit{D}}_{\mathit{L1}}}{\underline{D}_{\mathit{R2}}}{\overline{\underline{N}}_{\mathit{L1}}}{\overline{\underline{N}}_{\mathit{L1}}}\end{eqnarray}
correspond to the \(\Delta b=2\) operators
\(({\overline{d}_{L}}{b_{R}})^{2}\) and
\(({\overline{b}_{L}}{d_{R}})^{2}\). These should similar in size to the
\(\Delta S=2\) operators, in which case the CP violation in the \(b\) sector is 3
or 4 orders of magnitude smaller than in the standard model. We have
recovered a classic superweak model.

The following 6-fermion operators can also feed CP violation into the quark
masses, and thus into \(\overline{\theta }\).
\begin{eqnarray}&&{\overline{\mathit{D}}_{\mathit{L2}}}{\mathit{D}_{\mathit{R2}}
}{\overline{\underline{E}}_{\mathit{L1}}}{\underline{E}_{\mathit{R1}}}{\overline{\underline{N}}_{\mathit{L1}}}{\overline{\underline{N}}_{\mathit{L1}}}\\&&{\overline{\underline{D}}_{\mathit{L2}}}{\underline{D}_{\mathit{R2}}}{\overline{\underline{E}}_{\mathit{L1}}}{\underline{E}_{\mathit{R1}}}{\overline{\underline{N}}_{\mathit{L1}}}{\overline{\underline{N}}_{\mathit{L1}}}\\&&{\overline{\mathit{D}}_{\mathit{Li}}}{\underline{D}_{\mathit{Rj
}}}{\varepsilon
_{\mathit{ij}}}{\overline{\underline{E}}_{\mathit{L1}}}{\underline{E}_{\mathit{R1}}}{\overline{\underline{N}}_{\mathit{L1}}}{\overline{\underline{N}}_{\mathit{L1}}}\end{eqnarray}
Thus the CP violating parts of quark masses can be of similar magnitude to
the neutrino masses. This by itself does not sufficiently suppress
\(\overline{\theta }\), but the detailed structure of the quark mass matrices can
lead to further suppression.

\section{Conclusion} There can be many other effects of the new flavor
physics, through nonrenormalizable effects and in particular through the
effects of the
\(X\) boson. For example we can expect anomalous couplings of standard
model gauge bosons to the third family. Flavor changing effects may surface
in \(B\)--\(\overline{B}\) and \(\mathit{D}\)--\(\overline{\mathit{D}}\)
mixing, with the result that the \(B\) factories may uncover flavor changing
effects rather than CP violation.

In conclusion, we have bypassed the usual approaches to electroweak
symmetry breaking and proceeded straight to the flavor problem. We have
suggested that there is a dynamically broken flavor gauge symmetry around
100 to 1000 TeV which generates a wide variety of multi-fermion operators.
Close to a TeV the remnant flavor symmetry breaks, fourth family masses
arise, and electroweak symmetry breaking occurs. We have explored the
interplay between quark and lepton sectors in the generation of mass matrices.
We have also seen how the suppression of CP violation in the quark sector is
similar to the suppression of neutrino masses. One of the first signals of this
picture could be the absence of CP violation at B factories.

\section*{Acknowledgement} This research was supported in part by the
Natural Sciences and Engineering Research Council of Canada. I thank the
organizers of this workshop for their support, and the KEK theory group
where this report was prepared.

\newpage\pagestyle{empty}
\begin{center} \includegraphics{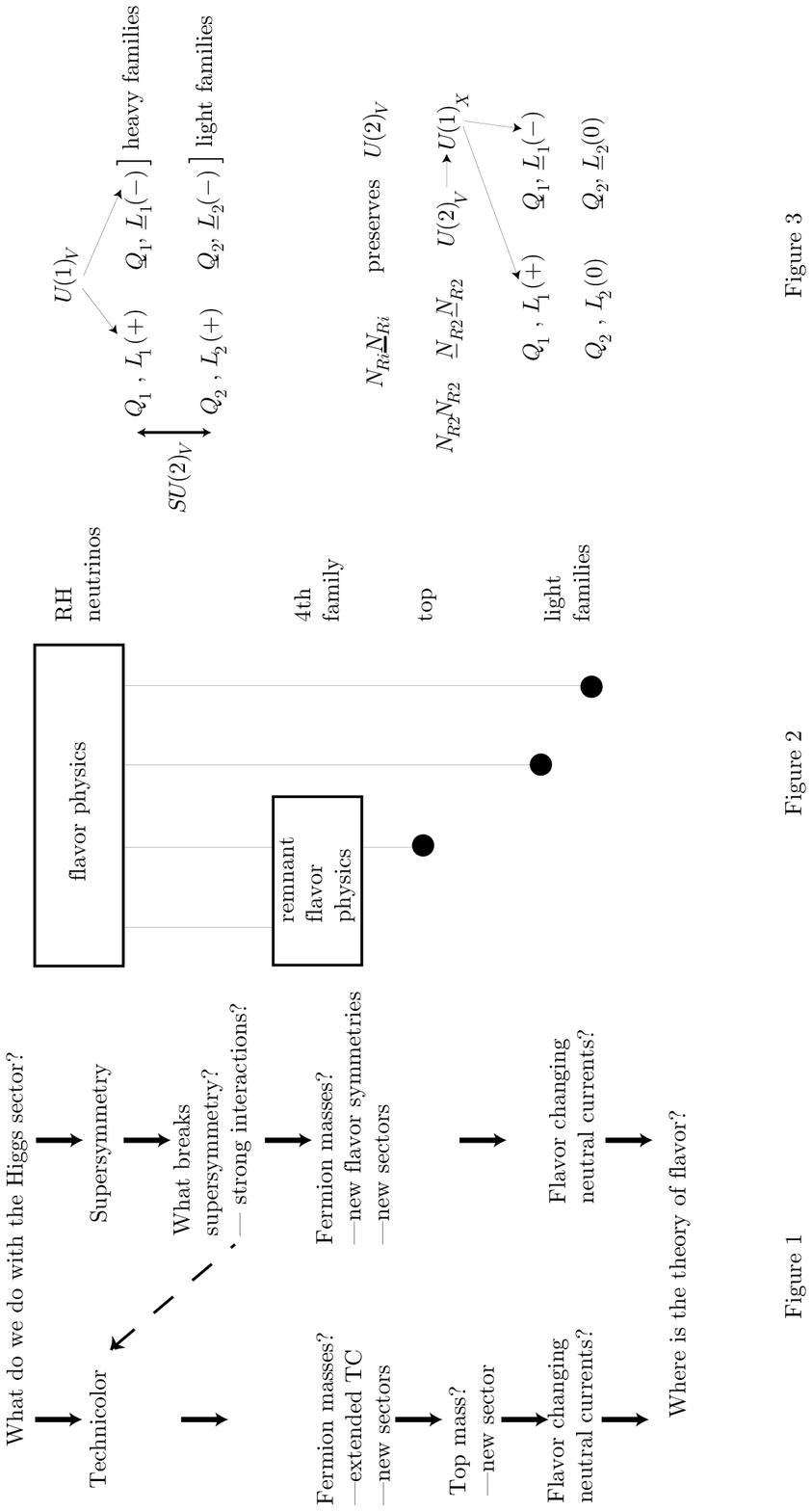}
\end{center}

\end{document}